\documentclass[iop]{emulateapj-rtx4}

\shorttitle{Swift reveals $\sim$5.7 day super-orbital period in XB158}
\shortauthors{Barnard et al.}

\begin{document}


\title{Swift reveals a $\sim$5.7 day super-orbital period in the M31 globular cluster X-ray binary  XB158}


\author{R. Barnard$^{1}$, M.~R. Garcia$^{1}$,  and S.~S. Murray$^{1,2}$}
\affil{$^1$Harvard-Smithsonian Center for Astrophysics, Cambridge, MA 02138, USA}

\affil{$^2$ Johns Hopkins University, Baltimore, MD, USA}


\begin{abstract}
The M31 globular cluster X-ray binary XB158 (a.k.a. Bo 158) exhibits  intensity dips on a 2.78 hr period in some observations, but not others. The short period suggests a low mass ratio, and an asymmetric, precessing disk due to additional tidal torques from the donor star since the disk crosses  the 3:1 resonance. Previous theoretical 3D smoothed particle hydrodynamical modeling suggested a super-orbital disk precession period 29$\pm$1 times the orbital period, i.e. $\sim$81$\pm$3 hr.  We conducted a Swift monitoring campaign of 30 observations over $\sim$1 month in order to search for evidence of such a super-orbital period. Fitting the 0.3--10 keV Swift XRT luminosity lightcurve with a sinusoid yielded a period of 5.65$\pm$0.05 days, and a $>$5$\sigma$ improvement in $\chi^2$ over the best  fit constant intensity model. A Lomb-Scargle periodogram revealed that periods 5.4--5.8 days were detected at a $>$3$\sigma$ level, with a peak at 5.6 days.  We consider this strong evidence for a 5.65 day super-orbital period, $\sim$70\% longer than the predicted period. The 0.3--10 keV luminosity varied by a factor $\sim$5, consistent with variations seen in long-term monitoring from Chandra. We conclude that other X-ray binaries exhibiting similar long-term behaviour are likely to also be X-ray binaries with low mass ratios and super-orbital periods.  
\end{abstract}





\section{Introduction}

The M31 globular cluster (GC) B158 (a.k.a. Bo 158), named following the Revised Bologna Catalogue v3.4 \citep{galleti04,galleti06,galleti07,galleti09}, contains a bright X-ray source that was discovered by the Einstein observatory \citep{tf91}, and has shown up in every X-ray observation of the region since; we call this X-ray source XB158. 

XB158 exhibited strong intensity modulation on a 10017$\pm$50 s ($\sim$2.78 hr) period during the 2002 January XMM-Newton observation \citep{trud02}. \citet{trud02} found similar variation in the folded lightcurves from  a 1991  June ROSAT observation and a 2000 June XMM-Newton observation; they found that the amplitude of modulation decreased with increasing source intensity. They found no such modulation in the 2001 June XMM-Newton observation, setting a 2$\sigma$ upper limit of 10\% modulation.  Assuming that this represents the orbital period, \citet{trud02} found that this is probably a  neutron star binary with a low mass donor with a separation $<$10$^{11}$ cm (i.e. a low mass XB-ray binary, LMXB). 

However, analysis of the unfolded 2000 June XMM-Newton lightcurve revealed a single deep dip at the end of the observation, with no evidence for dips in the two previous orbital cycles \citep{barnard06}. Furthermore, \citet{barnard06} analyzed 3 proprietary XMM-Newton observations over 2004 July 17--19, finding $\sim$100\% dipping for one orbital cycle, and zero evidence for dips in other cycles; we concluded that the disk was precessing, with dips only visible for some part of the super-orbital cycle.  

Such behavior is associated with the ``superhump'' phenomenon that is observed in accreting binaries where the mass ratio is smaller than $\sim$0.3 \citep{wk91}. Superhumps were first identified in the superoutbursts of the SU UMa subclass of cataclysmic variables (accreting  white dwarf binaries), manifesting as a periodic increase in the optical brightness on a period that is slightly longer than the orbital period \citep{vogt74,warner75}. SU UMas are a subclass of dwarf novae with orbital periods $\la$2 hr, that exhibit occasional superoutbursts that last $\ga$5 times as long as the normal outbursts \citep{vogt80}.

\citet{osaki89} proposed that these superoutbursts are enhanced by a tidal instability that occurs when the outer disk crosses the 3:1 resonance with the secondary; the additional tidal torque causes the disk to elongate and precess, and also greatly enhances the loss of angular momentum (and therefore the accretion rate). The disk precession is prograde in the rest frame, and the secondary repeats its motion with respect to the disk on the beat period between the orbital period and the precession period, a few percent longer than the orbital period. The secondary modulates the disk's viscous dissipation on this period, giving rise to the maxima in the optical lightcurve known as superhumps. Some short period, persistently bright CVs exhibit permanent superhumps \citep{patterson99,retter00}.

\begin{table*}
\begin{center}
\caption{Journal of Swift observations. For each observation we give the time (in days) after the first observation, pointing position for the observation, XRT  exposure time, and off-axis angle for XB158. We also give the net source counts from the observation} \label{journ}
\begin{tabular}{cccccccccccc}
\tableline\tableline
Obs & $T-T_0$ &  RA & Dec & XRT EXP & $D$/$'$ & Counts \\ 
\tableline \\
00032702001  & 0.0 &  00 43 00.81 & +41 17 00.2 & 2234 & 10.0 & 57 \\
00032702002& 0.9 &  00 43 06.53 & +41 16 54.6 & 2510 & 9.7 & 41 \\
00032702003 & 2.0 &  00 43 03.73 & +41 16 40.8 & 2299 & 9.5 & 13 \\
00032702004 & 3.0 &  00 43 06.35 & +41 16 51.3 & 2299 & 9.6 & 40 \\
00032702005 & 3.1 &  00 43 13.10 & +41 16 30.4 & 2489 & 9.2 & 43  \\
00032702006 & 4.4 &  00 43 03.22 & +41 13 58.5 & 2572 & 7.0 &  65\\
00032702007 & 5.4 &  00 43 10.44 & +41 15 10.9 & 2511 & 7.9 & 51  \\
00032702008 & 6.4 &  00 43 06.44 & +41 16 06.4 & 2635 & 8.9 & 57 \\
00032702009 & 7.3 &  00 43 11.97 & +41 16 27.7 & 2476 & 9.1 & 29   \\
00032702010 & 8.8 &  00 43 03.76 & +41 16 15.5 & 2644 & 9.1 & 39  \\
00032702011 & 9.5 &  00 43 08.40 & +41 17 27.6 & 2405 & 10.2 & 53  \\
00032702012 & 10.1 &  00 43 07.20 & +41 16 29.8 & 2502 & 9.2 & 43  \\
00032702013 & 11.2 &  00 43 03.63 & +41 17 36.2 & 2514 & 10.5 & 48   \\
00032702014 & 12.4 &  00 43 04.82 & +41 17 00.3 & 2411 & 9.8 &  38 \\
00032702015 & 13.2 &  00 43 07.68 & +41 15 24.0 & 2490 & 8.2 &  11 \\
00032702016 & 14.2 &  00 43 01.12 & +41 15 11.3 & 2511 & 8.2 & 27  \\
00032702017 & 15.1 &  00 43 06.35 & +41 14 46.6 & 2273 & 7.6 & 54  \\
00032702018 & 16.8 &  00 43 02.79 & +41 17 00.9 & 2314 & 9.9 & 55 \\
00032702019 & 17.1 &  00 43 08.59 & +41 15 26.4 & 1992 & 8.2 & 45 \\
00032702020 & 18.1 &  00 43 04.88 & +41 17 31.5 & 1389 & 10.3 & 13 \\
00032702021 & 19.4 &  00 43 03.24 & +41 17 28.0 & 2690 & 10.3 & 18 \\
00032702022 & 20.2 &  00 43 02.68 & +41 17 26.6 & 2686 & 10.3 &  42\\
00032702023 & 21.1 &  00 43 03.07 & +41 17 07.0 & 2821 & 10.0 & 65\\
00032702024 & 22.3 &  00 43 04.56 & +41 16 19.5 & 2659 & 9.2 & 50 \\
00032702025 & 23.2 &  00 43 10.49 & +41 16 54.1 & 2711 & 9.6 & 46  \\
00032702026 & 24.3 &  00 43 07.86 & +41 15 05.1 & 2740 & 7.8 & 35  \\
00032702027 & 25.1 &  00 43 06.42 & +41 16 41.0 & 2424 & 9.5 & 18\\
00032702028 & 26.1 &  00 43 10.04 & +41 16 54.2 & 2550 & 9.6  &49  \\
00032702029 & 27.2 &  00 43 10.52 & +41 17 33.9 & 2524 & 10.2 & 59 \\
00032702030 & 28.4 &  00 43 06.50 & +41 17 25.9 & 2331 & 10.2 & 41 \\
\tableline
\end{tabular}

\end{center}
\end{table*}

4U 1916$-$053 is a high inclination neutron star LMXB with an X-ray period of 50.00$\pm$0.08 min and an optical period 50.458$\pm$0.003 min \citep{callanan95}. It exhibits periodic X-ray intensity dips due to absorption by material in the outer disk; the amplitude of these dips varies over $\sim$0 to $\sim$100\% on a $\sim$4 day period, the precession period of the disk \citep{church98,chou01}. \citet{haswell01} showed that NS LMXBs with orbital periods shorter than $\sim$4.2 hr are likely to exhibit superhumps, and identified 4U 1916-053 as a persistent superhumping source. XB158 appears to be somewhat analogous to 4U 1916$-$053 \citep{barnard06}.

In \citet{barnard06} we modeled the 2004 July 17 XMM-Newton pn spectrum of XB158 with a blackbody and a power law, finding $kT$ = 2.0$\pm$0.2 keV, the photon index $\Gamma$ = 2.0$\pm$0.3, $\chi^2$/dof = 19/19 and the 0.3--10 keV luminosity was 1.5$\pm$0.6$\times 10^{38}$ erg s$^{-1}$; fitting a single power law emission model yielded a photon index of 0.57$\pm$0.09, which is harder than any black hole spectrum \citep{remillard06}, meaning that the accretor is a neutron star. We conducted  three dimensional smoothed particle hydrodynamical (SPH) modeling, assuming a 1.4 $M_\odot$ NS, a 2.78 hr period, a total system mass of 1.8 $M_\odot$, and  a luminosity $\sim$80\% of the Eddington limit. As a result, we  estimated the disk precession period to be 29$\pm$1 times the orbital period, i.e. 81$\pm$3 hr.

In \citet{barnard2012c} we examined  Chandra observations of $\sim$30 M31 globular cluster X-ray sources  spanning $\sim$12 years (89 ACIS and 45 HRC), including XB158. We observed 0.3--10 keV luminosities $\sim$4--20$\times 10^{37}$ erg s$^{-1}$, and proposed that this is due to varying accretion rates over the disk precession cycle.

Since we expected a super-orbital period for XB158 over time-scales of a few days, we obtained a series of 30 Swift observations, each with 2.5 ks exposure and spaced $\sim$1 day apart. In this work we present our analysis of these Swift results, and find evidence for a super-orbital period of $\sim$6 days.

\section{Observations and data analysis}

We made 30 Swift observations over 2013 February 8 -- March 9 (ObsIDs 0032702001--0032702030, PI R. Barnard). XB158 was one of two main targets for this survey, and the pointing was chosen to optimize the results from both targets. However, the actual pointings varied significantly over the 30 observations. A journal of Swift observations is provided in Table~\ref{journ}; for each observation we give the time relative to the first observation, pointing , XRT exposure, and off-axis angle, along with the net source counts.

For each observation, we placed circular  regions around XB158 and a suitable background region. We obtained the  net source counts using the ``Counts in regions'' tool in the DS9 image viewer, and estimated the intensity by dividing the net counts by the exposure time. These data were obtained in order to determine the extent to which  the varying off-axis angles affected our results.

We also created spectra from the same extraction regions, created appropriate ancillary response files using XRTMKARF, and found the appropriate response file using the QUZCIF tool.
None of the spectra were suitable for free  spectral fitting, hence we obtained luminosity estimates by assuming  a model obtained from previous observations. For Chandra ACIS observations with $>$200 net source photons, the mean line-of-sight absorption ($N_{\rm H}$) was  9$\pm$4$\times 10^{20}$ atom cm$^{-2}$ ($\chi^2$/dof = 7/11), and the mean power law index ($\Gamma$) was  0.52$\pm$0.03 ($\chi^2$/dof = 5/11).  This is consistent with the best absorbed power law fit to the 2004 July 17 XMM-Newton observation \citep[$N_{\rm H}$ = 0.1, $\Gamma$ = 0.57$\pm$0.09, $\chi^2$/dof = 30/24][]{barnard06}. As we noted earlier, that XMM-Newton spectrum was best described by a blackbody + power law model, but neither the Chandra nor Swift spectra were sufficient to constrain the two-component emission model.

However, we were able to estimate the luminosity for each observation by assuming a fixed emission model, allowing only the normalization to vary. We fitted each spectrum using XSPEC 12.8.2b,  fixing  $N_{\rm H}$ = 9$\times 10^{20}$ atom cm$^{-2}$ and $\Gamma$ = 0.52, to find the normalization required to make absorbed model intensity 1.00 count s$^{-1}$. We then calculated the unabsorbed flux for this model, allowing us to convert from intensity to flux. Multiplying the conversion by the background-subtracted  intensity provided by XSPEC yielded the instrument-corrected, background-subtracted source flux.  The luminosity was calculated from the flux assuming a distance of 780 kpc \citep{stanek98}.

We fitted the lightcurve with constant and sinusoidal components using the QDP program provided in HEATOOLS, performing a simple search for periodicity. \citet{scargle82} created a periodogram suitable for unbinned data with a mean of zero that produces exactly equivalent results to such least-squares fitting of sinewaves, but also allows comparison of the best period with other periods. 
 The likelihood that any peak in the periodogram is real is given by the false alarm probability ($P$), where a low value of $P$ indicates that the peak is likely to be  significant. If the highest frequency to be probed is $N$ times higher than the lowest frequency, then the power, $z$, required for a false alarm probability $P$ is given by $z$ = $-\ln \left[1-\left(1-p\right)^{1/N}\right]$; for small $P$, $z \sim \ln\left(N/P\right)$ \citep{scargle82}.
\section{Results}
\label{res}


\begin{figure}
\epsscale{1.2}
\plotone{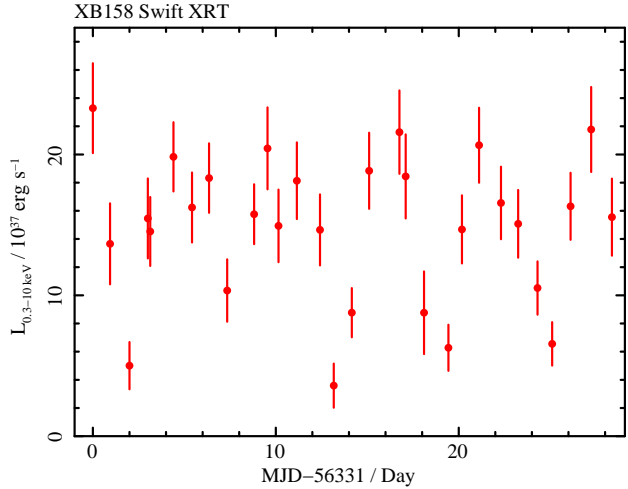}
\caption{ 0.3-10 keV Swift luminosity lightcurve of XB158 for our 30 days of observations.  }\label{swiftlc}
\end{figure}

We present our 30 day,  0.3--10 keV Swift XRT lightcurve of XB158 in Figure~\ref{swiftlc}. We see that the 0.3--10 keV luminosity varied by a factor $\sim$5; the luminosity dropped from $\sim$2.3$\times 10^{38}$ erg s$^{-1}$ to $\sim$5$\times 10^{37}$ erg s$^{-1}$ in $\sim$2 days, assuming the mean Chandra absorbed power law model. We note that these Swift observations are non-contiguous, spacing the 2.5 ks observing time over several hours; the low intensities ($\sim$0.005--0.025 count s$^{-1}$) meant that there was no appreciable variability within each observation.

 We find no evidence for a dependence of luminosity on off-axis angle; the luminosities for the observations when XB158 have the largest and smallest off-axis angles have consistent values, while observations at an off-axis angle of 8.2$'$ resulted in a factor $\sim$5 range in luminosity. The conversion factor for translating 1 count s$^{-1}$ into 0.3--10 keV flux ranged over 1.09--1.29$\times 10^{-10}$ erg cm$^{-2}$ count$^{-1}$ for all observations except the second one, where it was 1.79$\times 10^{-10}$ erg cm$^{-2}$ count$^{-1}$. This $\sim$10\% variation about the mean in instrumental correction is clearly not sufficient to account for the factor $\sim$5 variation in luminosity.

Fitting our lightcurve with a constant intensity yielded a best fit luminosity of $\sim$1.3$\times 10^{38}$ erg s$^{-1}$, with $\chi^2$ = 181 for 29 degrees of freedom (dof). However, adding a sinusoidal variation component yielded a much improved fit, with $\chi^2$/dof = 43/26; for this model, the period is 5.65$\pm$0.05 days, with an amplitude of 7.1$\pm$0.6$\times 10^{37}$ erg s$^{-1}$ around a mean luminosity of 1.43$\pm$0.04$\times 10^{38}$ erg s$^{-1}$, and a phase of 88.1$\pm$0.7 degrees. All uncertainties in this work are quoted at the 1$\sigma$ level.

\begin{figure}
\epsscale{1.2}
\plotone{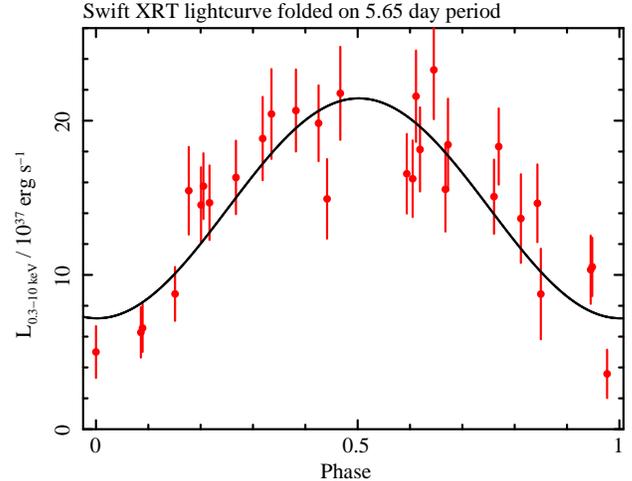}
\caption{ Folded 0.3--10 keV Swift XRT luminosity lightcurve, assuming a 5.65 day  super-orbital period.}\label{swift_per}
\end{figure}

This sinusoidal variability yielded $\Delta\chi^2$ = 138 for 30 bins, with 3 extra free parameters; F-testing showed that the probability for this improvement being due to chance was 3$\times 10^{-8}$, equivalent to a $>$5$\sigma$  detection. In Figure~\ref{swift_per} we show our Swift lightcurve folded on a 5.65 day period and fitted with the best fit sinusoid.

\begin{figure}
\epsscale{1.2}
\plotone{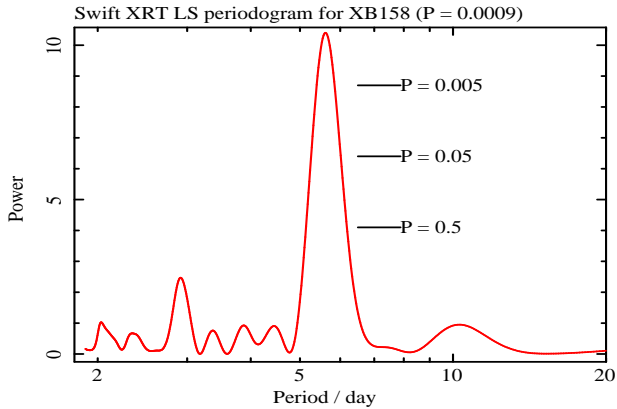}
\caption{ Lomb-Scargle periodogram for our 0.3--10 keV Swift XRT luminosity lightcurve, identifying the power required for false alarm probabilites ($P$) = 0.5, 0.05, and 0.005. We see a peak at 5.6 days with $P$ = 0.0009, while 5.4--5.8 day periods are significant at a $>$3$\sigma$ level.}\label{ls}
\end{figure}

We present our Lomb-Scargle periodogram for the 0.3--10 keV Swift XRT luminosity lightcurve of XB158 in Figure~\ref{ls}, and also indicate the power required for false alarm probabilities $P$ = 0.5, 0.05, and 0.005 for reference. We tested 30 frequencies,  and oversampled each frequency by a factor of 50; this resulted in a periodogram covering a wider range of periods than is interesting (up to $\sim$1500 days), so we show only part of the periodogram here.  The periodogram shows a single strong peak, with the maximum power of 10.4 at a period of 5.60 days, corresponding to $P$ = 0.0009; the range of periods which are detected at a $>$3$\sigma$ level is 5.4--5.8 days. While there is a small peak at the 3 day period, $P$ = 1 for this peak.


 \section{Discussion and Conclusions}
XB158 is a high inclination X-ray binary associated with the M31 globular cluster B158. It exhibits deep intensity dips on a 2.78 hr period in some observations but not others, prompting \citet{barnard06} to suggest that the disk is precessing, caused by the ``superhumping'' phenomenon observed in low mass ratio systems where the disk crosses the 3:1 resonance with the donor star. 

\citet{barnard06} predicted a disk precession period of 29$\pm$1 times the orbital period, i.e. 81$\pm$3 hr, inspiring a month of daily monitoring of the M31 central region by Swift. Fitting a sinusoid to the lightcurve  revealed a 5.65$\pm$0.05 day super-orbital period (1$\sigma$ uncertainties);  the 0.3--10 keV luminosity varied by a factor $\sim$5, which is consistent with the range of luminosities observed in the ACIS observations of our 13+ year Chandra monitoring campaign \citep{barnard2012c}.

The peak of the Lomb-Scargle periodogram is at 5.6 days,  consistent with that obtained from least squares fitting of the lightcurve with a sinusoid.
The suggested super-orbital period is $\sim$70\% longer than predicted by our 3D SPH simulations \citep{barnard06}. None of the authors of the current paper are experts in SPH; however, J.~R. Murray stated in a private communication that the longer than expected super-orbital period is likely due to the mass ratio of the donor to the accretor being lower than assumed. The mass of a Roche lobe filling main sequence star, $m$, may be approximated as $m$ $\simeq$ 0.11 $P_{\rm hr}$, where $P_{\rm hr}$ is the orbital period in hours \citep{frank02}, and we originally assumed a donor mass of $\sim$0.30 $M_\odot$. For the accretor, we assumed a 1.4 $M_{\odot}$ neutron star. A power law emission model fit to the 2004 July 16 XMM-Newton spectrum of B158 yielded a photon index of 0.57$\pm$0.09 \citep{barnard06}, which is considerably harder than any spectrum emitted by a black hole binary. However, some neutron stars have masses $>$2$M_\odot$ \citep[see e.g.][]{demorest10,lynch13}, and it is possible that XB158 contains a particularly massive neutron star.

We noted in \citet{barnard2012c} that other XBs such as XB146 exhibited strong luminosity fluctuations between fairly consistent maxima and minima during our Chandra monitoring observations, and suggested that this long-term behavior could be indicative of a short period  / low mass ratio system. Our new findings support this hypothesis.

\section*{Acknowledgments}
We thank the anonymous referees for suggesting improvements to this work, in particular prompting more rigorous estimation of the super-orbital period. We thank the Swift team for making this work possible. 
This work was funded by Swift grant NNX13AJ76G.





{\it Facilities:} \facility{Swift (XRT)} 






\clearpage



\end{document}